\documentclass[%
 reprint,
 amsmath,amssymb,
 aps,
]{revtex4-2}

\usepackage{graphicx}
\usepackage{dcolumn}
\usepackage{bm}
\usepackage{ragged2e}
\usepackage[justification=raggedright,format=plain]{caption}
\usepackage{subcaption}
\usepackage{fbb}

\begin{document}


\title{Effect of Dormant Spare Capacity on the Attack Tolerance of Complex Networks}

\author{Sai Saranga Das M}
 \affiliation{Department of Biotechnology, Bhupat and Jyoti Mehta School of Biosciences, IIT Madras}
\author{Karthik Raman}
 \email{kraman@iitm.ac.in}
\affiliation{Department of Biotechnology, Bhupat and Jyoti Mehta School of Biosciences, IIT Madras}
\affiliation{Robert Bosch Centre for Data Science and Artificial Intelligence (RBCDSAI), IIT Madras}
\affiliation{Centre for Integrative Biology and Systems Medicine (IBSE), IIT Madras}

\begin{abstract}
The vulnerability of networks to targeted attacks is an issue of widespread interest for policymakers, military strategists, network engineers and systems biologists alike. Current approaches to circumvent targeted attacks seek to increase the robustness of a network by changing the network structure in one way or the other, leading to a higher size of the largest connected component for a given fraction of nodes removed. In this work, we propose a strategy in which there is a pre-existing, dormant spare capacity already built into the network for an identified vulnerable node, such that the traffic of the disrupted node can be diverted to another pre-existing node/set of nodes in the network. Using our algorithm, the increase in robustness of canonical scale-free networks was nearly 16-fold. We also analysed real-world networks using our algorithm, where the mean increase in robustness was nearly five-fold. To our knowledge, these numbers are significantly higher than those hitherto reported in literature. The normalized cost of this spare capacity and its effect on the operational parameters of the network have also been discussed. Instances of spare capacity in biological networks, termed as distributed robustness, are also presented. 
\end{abstract}

\maketitle


\section{\label{sec:level1}Introduction}

Power-law networks, owing to their distinctive structure, are inherently safeguarded from random failures~\citep{Albert2000Error,Barabasi1999Emergence,caldarelli2007scale}. Yet, the very attribute that makes them extremely resistant to random failures also makes them easy targets for targeted attacks by agents with subversive interest~\citep{cohen2001breakdown,albert2004structural}. To increase the robustness of networks against such attacks, numerous strategies have been devised till date. These strategies are typically based on repositioning the existing edges of a network (Edge Swapping or ES strategy) or adding more edges to the network (Edge Addition or EA strategy). One of the most effective ways reported in the literature to increase the robustness of a network by the ES method is to give it an onion-like structure~\citep{Schneider2011Mitigation},  with a core of densely connected nodes hierarchically surrounded by layers of nodes of decreasing degree. Additionally, there have  been other works that seek to further increase the robustness of networks either by the ES~\citep{Louzada2013Smart,Krol2015Propagation,rong2018heuristic,wu2011onion,tanizawa2012robustness} or the EA strategy ~\citep{Li2019Maximizing,jiang2011enhancing,paul2004optimization,zhao2009enhancing}. Both the ES and the EA strategies are based on the smart rewiring of an existing network by swapping or adding edges in such a way that, in the event of an attack, the size of the largest connected component, $s(\mathbf{q})$,~\citep{Schneider2011Mitigation} is within a functionally acceptable range. 

The work of Schneider \textit{et al.}~\citep{Schneider2011Mitigation} that describes the onion-like network structure is based on a key assumption---the invariance of degree distribution. In other words, they assumed that increasing the degree of a node is more expensive than swapping the edges between node pairs. This assumption, implicit in all ES strategies, holds well when we change the degree of a node in the network from $n$ to $k$, where $k \gg n$. However, around the region of $n$, the cost of this change should be as much as the cost of swapping the edges, if not lower

In this study, we propose a novel EA algorithm for increasing network robustness. Our aim is to define a ``spare capacity'' of edges for a given network. In other words, in the event of a node's disruption, the dormant spare capacity embedded in the network will take over and reroute the traffic of the disrupted node, thereby maintaining its informational integrity. Further, our aim is to minimise the cost of this spare capacity, which is a factor of the number of edges that are added in the network and the increase/decrease in the path length of the modified network. Our algorithm also imposes a limit on the increase in the degree of a given node during the addition of spare capacity --- we call this the acceptable \emph{degree deviation} for a given node in the network. 

\section{Methods}
\label{sec:methods}

\subsection{Measures of Robustness and Spare Capacity}
The robustness of a network is measured as the size of the largest connected component post disruption of a specific set of nodes. In this regard, the measure used in~\citep{Schneider2011Mitigation} has been used in this study. Spare capacity is added to a given network as a percentage of the total number of edges originally present in it. 

\subsection{Generation of canonical networks}

Scale-free networks were generated using the preferential attachment algorithm described by~\citep{Barabasi1999Emergence}. These networks have a degree distribution given by $P(k) \propto k^{-\gamma}$, with $\gamma = 2.5$.
$G(n,p)$ Erdős--Rényi networks were also generated for analysis. The codes for the generation of scale-free networks, along with those for the application of our algorithm were developed in-house, and are available via GitHub (\url{https://github.com/SaiSarangaDas/Spare-Capacity-Algorithm/}). The code for the generation of Erdős--Rényi networks was taken from Matlab BGL package (\url{https://in.mathworks.com/matlabcentral/fileexchange/10922-matlabbgl}).

\subsection{Algorithm to compute the spare capacity for a given network}
\label{sub:algorithm}

\textbf{Input:} A network whose spare capacity has to be determined.

\textbf{Output:} The modified network with added spare capacity, the disrupted network,  cost of spare capacity for the network, the degree of nodes in the original network for which optimal spare capacity has been added.

\textbf{Step 1:} Perform a network disruption using a centrality measure of choice. In this study, betweenness centrality and degree centrality were used as a measure for node disruption.

\textbf{Step 2:} Identify the two largest connected components in the disrupted network. Let them be $\mathbf{a}$ and $\mathbf{b}$. The vectors $\mathbf{t}$ and $\mathbf{v}$ will contain the node numbers of the nodes present in $\mathbf{a}$ and $\mathbf{b}$. The reason for our focus on the two largest connected components of the network is simple---an edge (spare capacity) between the two largest connected components in the network will result in the least number of infinite path lengths per unit spare capacity, thereby increasing the size of the largest connected component of the network.

\textbf{Step 3:} For every \emph{potential} edge between pair of nodes in $\mathbf{a}$ and $\mathbf{b}$ respectively, evaluate the sum of finite path lengths of the network. The vector $\mathbf{q}$ will contain the node numbers and the sum of finite path lengths for all possible node pairs between $\mathbf{a}$ and $\mathbf{b}$.

\textbf{Step 4:} Sort $\mathbf{q}$ in the ascending order of finite path lengths. The first row in $\mathbf{q}$ is the optimal spare edge between the two connected components $\mathbf{a}$ and $\mathbf{b}$ since an edge between these two node pairs results in the least sum of finite path lengths in the network.

\textbf{Step 5:} If the degree of either node for which the spare capacity is to be added surpasses a certain threshold, choose the next best spare edge from $\mathbf{q}$.

\textbf{Step 6:} Perform steps 1 to 5 iteratively for the two successive largest connected components in the network so as to identify the full set of spare capacity for the same.

\subsection{Datasets}

All real-world networks investigated in this study, were obtained from Network Repository (\url{https://networkrepository.com/}; \cite{nr}).

\section{Results} \label{sec:results}

\begin{figure}[h]
 \includegraphics[width=\linewidth]{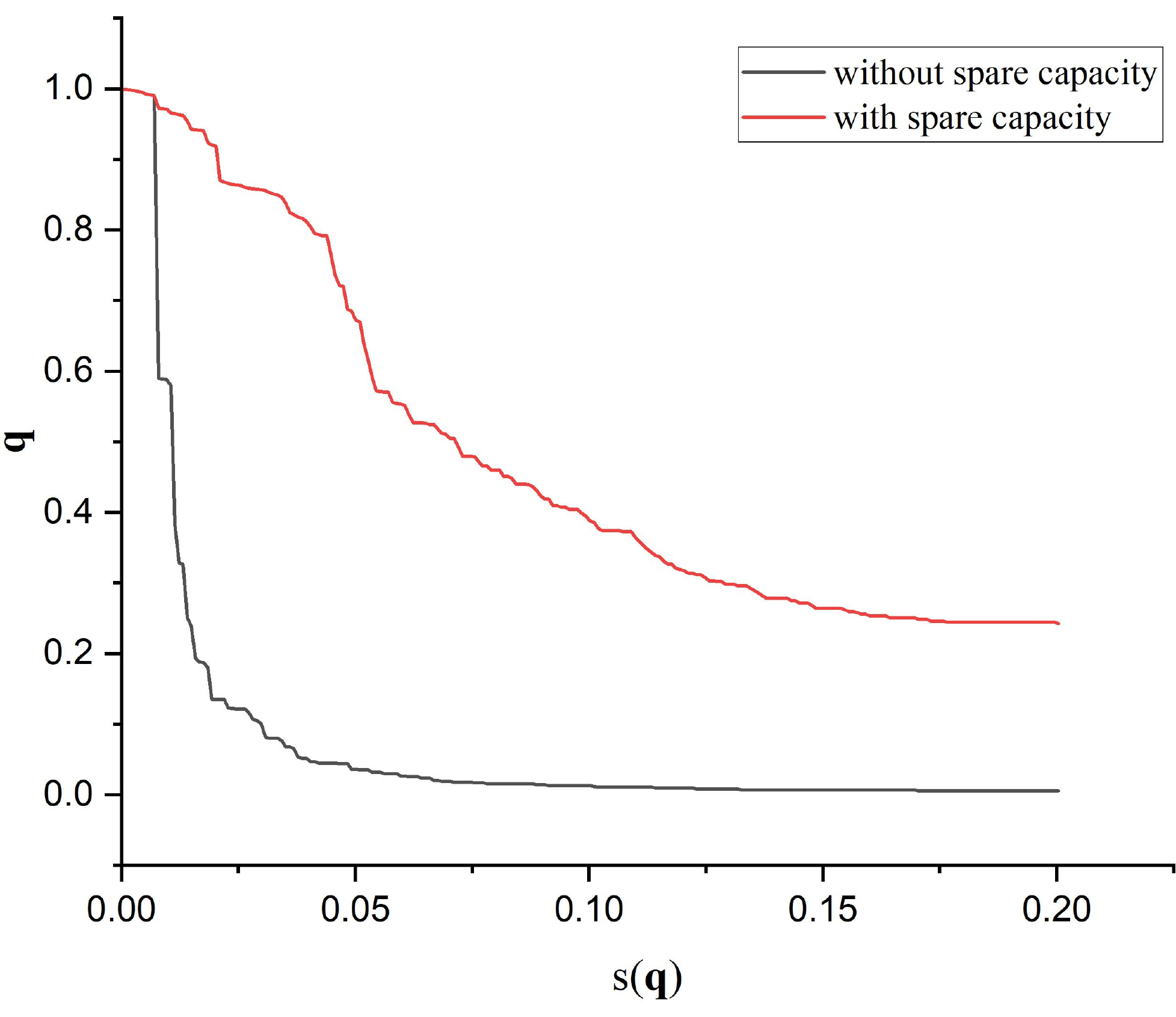}
 \caption{A typical Robustness graph where $s(\mathbf{q})$ is computed for a network without and with 5\% spare capacity (black and red plots respectively). The graph given above represents the robustness graph of 1138-bus network when betweenness centrality is used as the method for node disruption (more below).}
 \label{fig:robustness-1138} 
 \end{figure}

We demonstrate the utility of designing networks with spare capacity, illustrating how the robustness of such networks remain fairly high even during the onslaught of targeted attacks. We illustrate results on both canonical Erdős--Rényi and scale-free networks, as well as a variety of real-world networks. A typical robustness graph, plotting $s(\mathbf{q})$ (size of the largest connected component) against $\mathbf{q}$ (number of nodes removed as a fraction of the total number of nodes in the network) has the characteristic appearance shown in Figure~\ref{fig:robustness-1138}, where the inclusion of spare capacity is seen to have a significant positive impact on network robustness.

\subsection{Robustness of Canonical Networks}
\label{sub:robustness-of-characteristic-networks}
In this section, we discuss the application of our algorithm to canonical Erdős--Rényi $G(n,p)$ and scale-free networks.

\subsubsection{Erdős--Rényi Networks}
\label{sub:er-networks}

Typical Erdős--Rényi networks, in the context of usage of our algorithm, do not yield an appreciable increase in the robustness value. For instance, at 5\% spare capacity, the highest increase in the robustness value obtained from our simulations for a 1,000 node network with a $p$ value of 0.003 were 52.4\% and 15.4\% when betweenness centrality and degree centrality were used as the mode of network disruption respectively. (At lower $p$ values, the generated network has a sizable number of clusters, resulting in artificially inflated values of robustness increase (data not shown)). For $p$ values greater than 0.005, our algorithm seldom leads to any robustness increase.  These results are not surprising; Erdős--Rényi networks, owing to their random nature, are immune to targeted attacks and withstand the attack pressure with relative ease~\citep{Albert2000Error}. 

\subsubsection{Power Law Networks} \label{ssub:power-law-networks}

\begin{figure*}
\begin{subfigure}[b]{0.48\textwidth}
\centerline{\includegraphics[width=\linewidth]{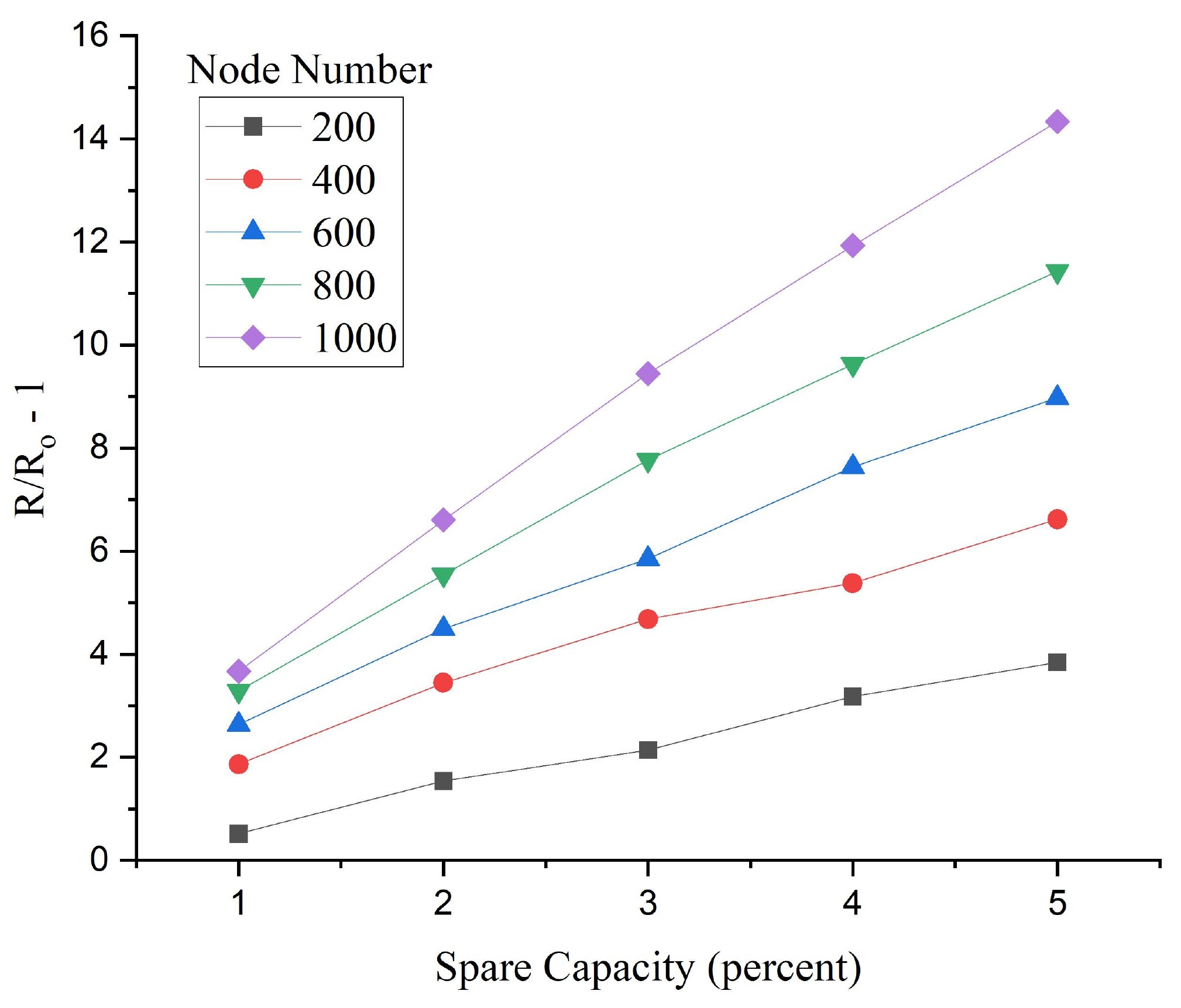}}
\caption{}
\label{fig:robustness-power-law-a}
\end{subfigure}\hfill
\begin{subfigure}[b]{0.48\textwidth}
\centerline{\includegraphics[width=\linewidth]{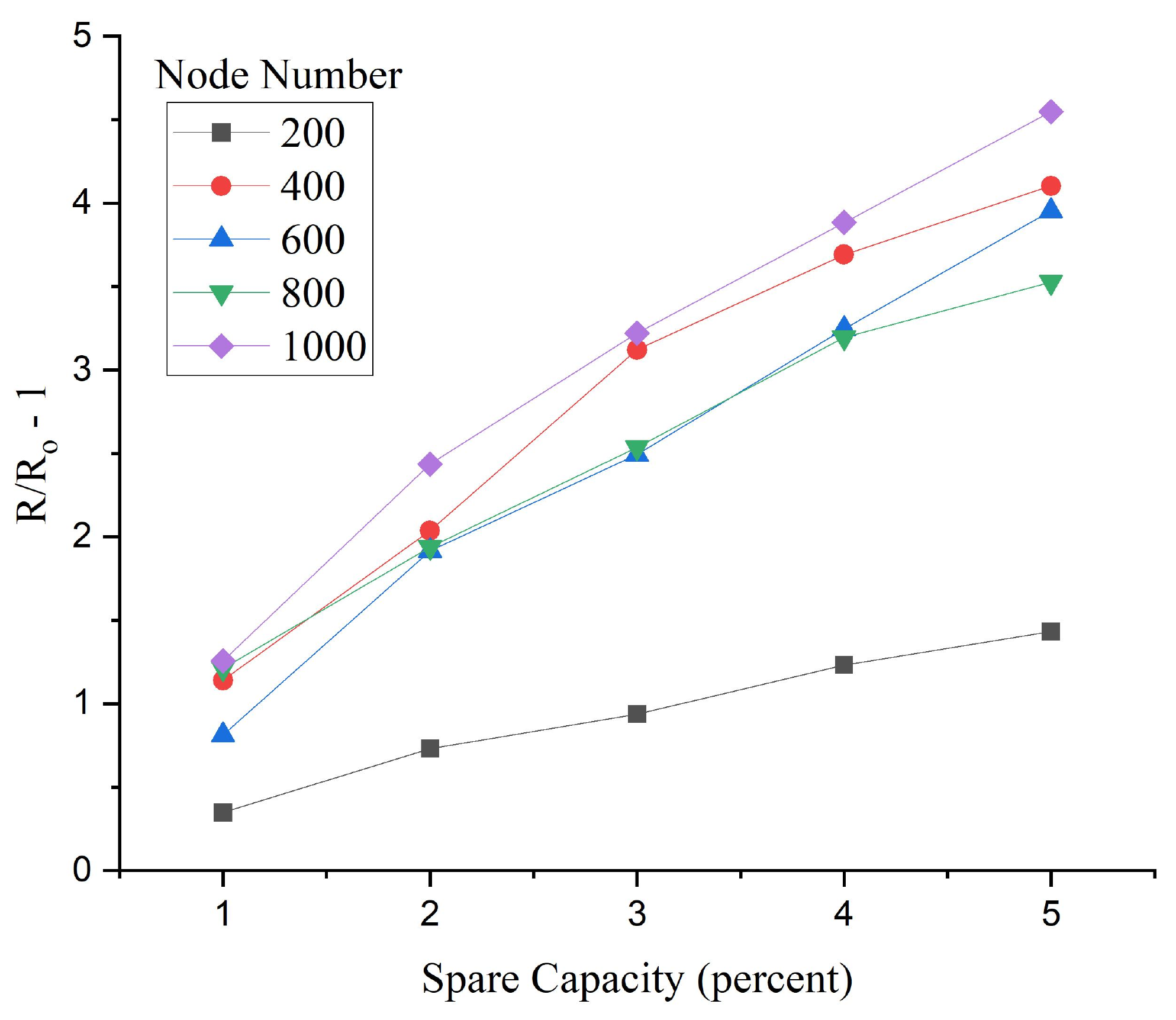}}
\caption{}
\label{fig:robustness-power-law-b}
\end{subfigure}
\caption{Robustness increase for 200 node (closed squares), 400 node (closed circles), 600 node (closed triangles), 800 node (closed inverted triangles) and 1,000 node (closed diamonds) power law networks with (a)~betweenness centrality and (b)~degree centrality being the method for node disruption. Gamma value for all networks was 2.5. The total number of nodes removed was 20\% of the original number of nodes in the network.}
\end{figure*}

As mentioned earlier, power law networks, owing to their inherent nature, are extremely vulnerable to targeted attacks. Consequently, the increase in robustness value of these types of networks are indicative of their fragility to targeted attacks, as shown in Figures~\ref{fig:robustness-power-law-a} and~\ref{fig:robustness-power-law-b}.

\begin{figure*}
\begin{subfigure}[b]{0.48\textwidth}
\includegraphics[width=\linewidth]{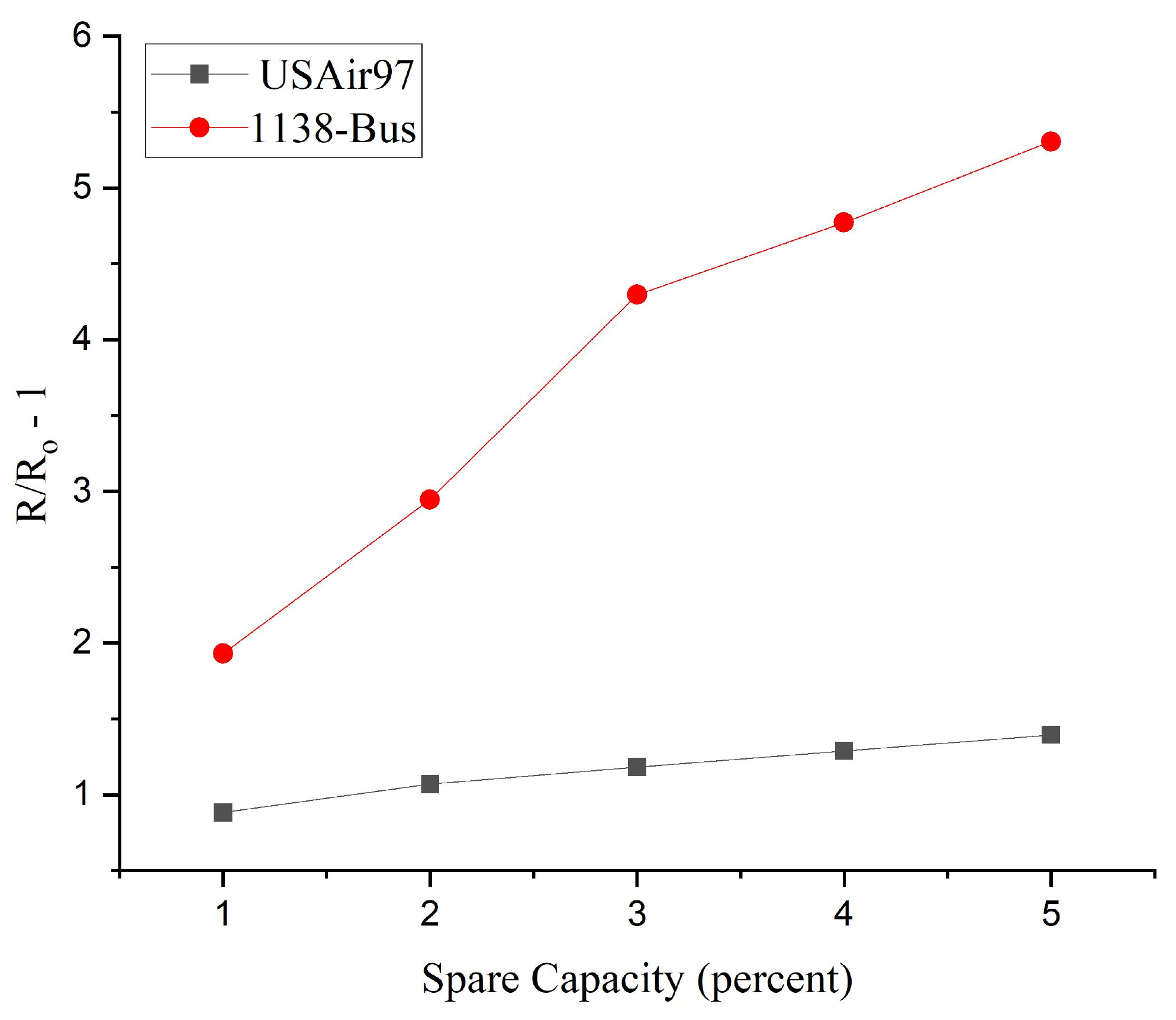}
\caption{}
\label{fig:betweenness-centrality-USAIR-1138-Bus}
\end{subfigure}\hfill
\begin{subfigure}[b]{0.48\textwidth}

\includegraphics[width=\linewidth]{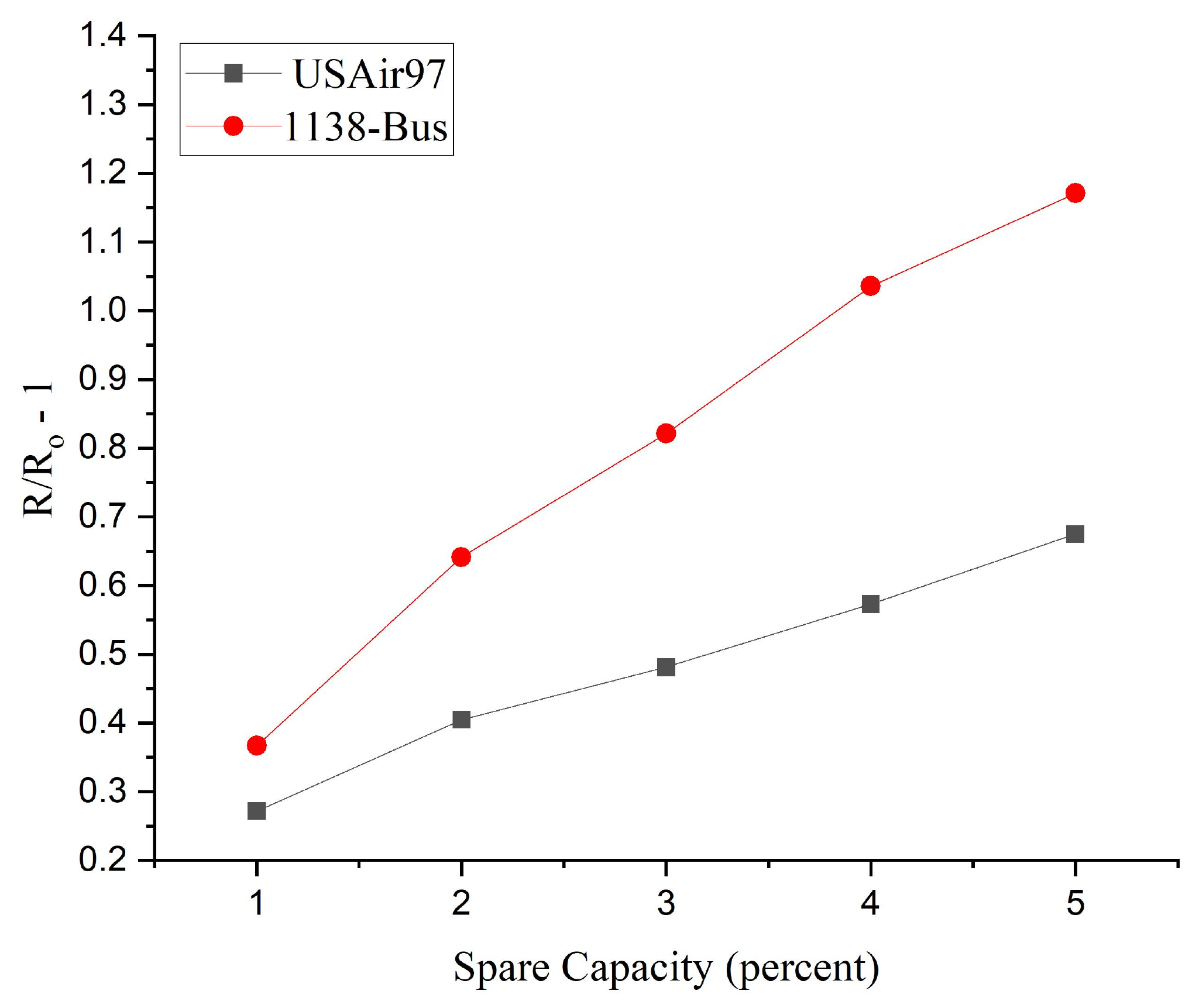}
\caption{}
\label{fig:degree-centrality-USAIR-1138-Bus}
\end{subfigure}
\caption{Robustness increase for USAir97 (closed squares) and 1138 Bus (closed circles).
Nodes removed based on (a)~betweenness centrality   and (b)~degree
centrality. Total number of nodes removed was 20\% of
the number of nodes in the network. Number of nodes and edges
in the networks: USAir97, 332 and 2,178 respectively; 1138-Bus, 1,138
and 1,458 respectively.}
\end{figure*}

A few key observations can be made from the figure. Firstly, for a network of any given size, the robustness increase for the disruption of nodes with betweenness centrality is higher than that for the disruption of nodes with degree centrality, an inference that reiterates the conclusion made in an earlier study ~\citep{holme2002attack}. Moreover, the standard deviation in the robustness increase of similar sized networks was higher when degree centrality was used as the mode of network disruption (data not shown).

Using our algorithm on a 1,000 node scale-free network, the improvement in the robustness value at 5\% spare capacity  was 1,434\% and 455\% when betweenness centrality and degree centrality was used as the measure for node disruption respectively. When compared with the results obtained from works on canonical scale free networks using both the ES and EA methods~\citep{Schneider2011Mitigation,Louzada2013Smart,wu2011onion,tanizawa2012robustness, rong2018heuristic,Li2019Maximizing}, our methodology shows a significant increase in the robustness value. 

\subsection{Example Real-world Networks}
\label{ssub:real-world-networks}

Two infrastructure networks---an air traffic network (USAir97) and a power distribution network (1138-bus)---were selected from Network Repository, a publicly available database of networks. The average degrees of USAir97 and 1138-bus networks were 6.6 and 1.3 respectively. As mentioned before, the degree of any given node, during the course of spare capacity addition, is not allowed to deviate over and above a given value. Numerically, this metric is measured as the \emph{degree deviation} of a node in a given network.

An interesting phenomenon from Figures~\ref{fig:betweenness-centrality-USAIR-1138-Bus} and ~\ref{fig:degree-centrality-USAIR-1138-Bus} is evident---the increase in the robustness value for 1138-bus is higher than that for USAir-97 regardless of the mode of network disruption. This is due to the fact that the decrease in robustness of the US air transport network (with and without spare capacity) is lower than that of 1138-bus (Figure~\ref{fig:robustness-1138} and Supplementary Figure~\ref{fig:robustness-us-air}). Since USAir97 with an average degree of 6.6 is more resistant to attacks than 1138-Bus with an average degree of 1.3, our initial understanding was that, owing to a preponderance of edges, USAir97 is relatively well connected to withstand targeted attack when compared to 1138-bus.

\begin{figure}
\includegraphics[width=\linewidth]{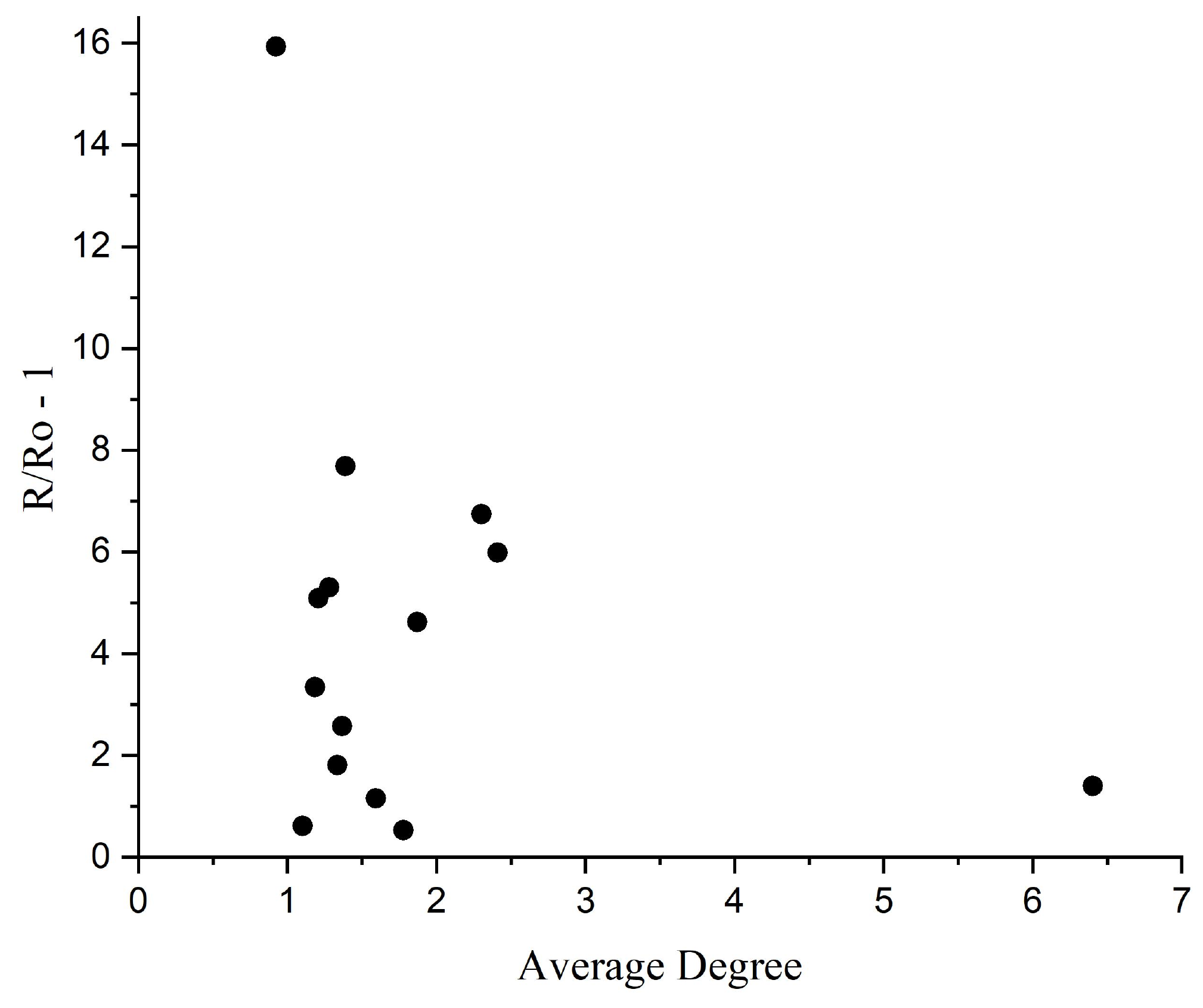}
\caption{The robustness increase at 5\% spare capacity of different real life networks analysed in this study when betweenness centrality was used as the measure of node disruption. Data is similar when degree centrality was used as the measure for node disruption (Supplementary Figure~S2). Number of nodes removed was 20\% of the total number of nodes in the network. For more information on the networks refer to Supplementary Table S1.}
\label{fig:different-networks-bc}
\end{figure}

To test our hypothesis, we analysed a sample of real-world networks (14 networks in total) to ascertain the nature of the relationship between the increase in robustness value and the average degree of networks (Figure~\ref{fig:different-networks-bc}). From the figure, it is evident that though the increase in robustness value is somewhat related to the average degree of a given network in a few instances, it cannot be generalised as a universal principle; each network has to be analysed individually based on its unique topology. However, in all but one instance, the added spare capacity decreases the characteristic path length of the network (Supplementary Table~S1). This is due to the nature of our algorithm and has major implications on the operational cost of the network when it operates with spare capacity.

\subsection{The Importance of Normalised Cost}
\label{sub:normalized-cost}

The most efficient way to add spare capacity to a network is by adding edges to its two largest connected components -- the methodology followed by our algorithm. Hence, regardless of an edge between any node pair between the two connected components, the increase in $s(\mathbf{q})$ of the network during the process of spare capacity addition is the same. However, the sum of finite path lengths of the nodes between the two connected components changes depending on which of the two nodes in the respective connected components is selected for spare capacity addition. An increase in this sum implies an increase in the characteristic path length of the network. This has some crucial implications. Generally, higher the sum, higher is the operational cost of the added spare capacity. As spare capacity is built successively into the entire network, the sum of sum of finite path lengths, or more aptly, the cost of spare capacity, becomes important in understanding the differences in the operational cost of the network when it is operating through activated spare capacity. In the case of the air transportation network, an increase in this metric implies an increase in the time and cost of transportation from one airport to another. In the case of the power network, the increase represents an increase in the duration of power transmission, with corresponding voltage losses. 

\begin{figure}
\includegraphics[width=\linewidth]{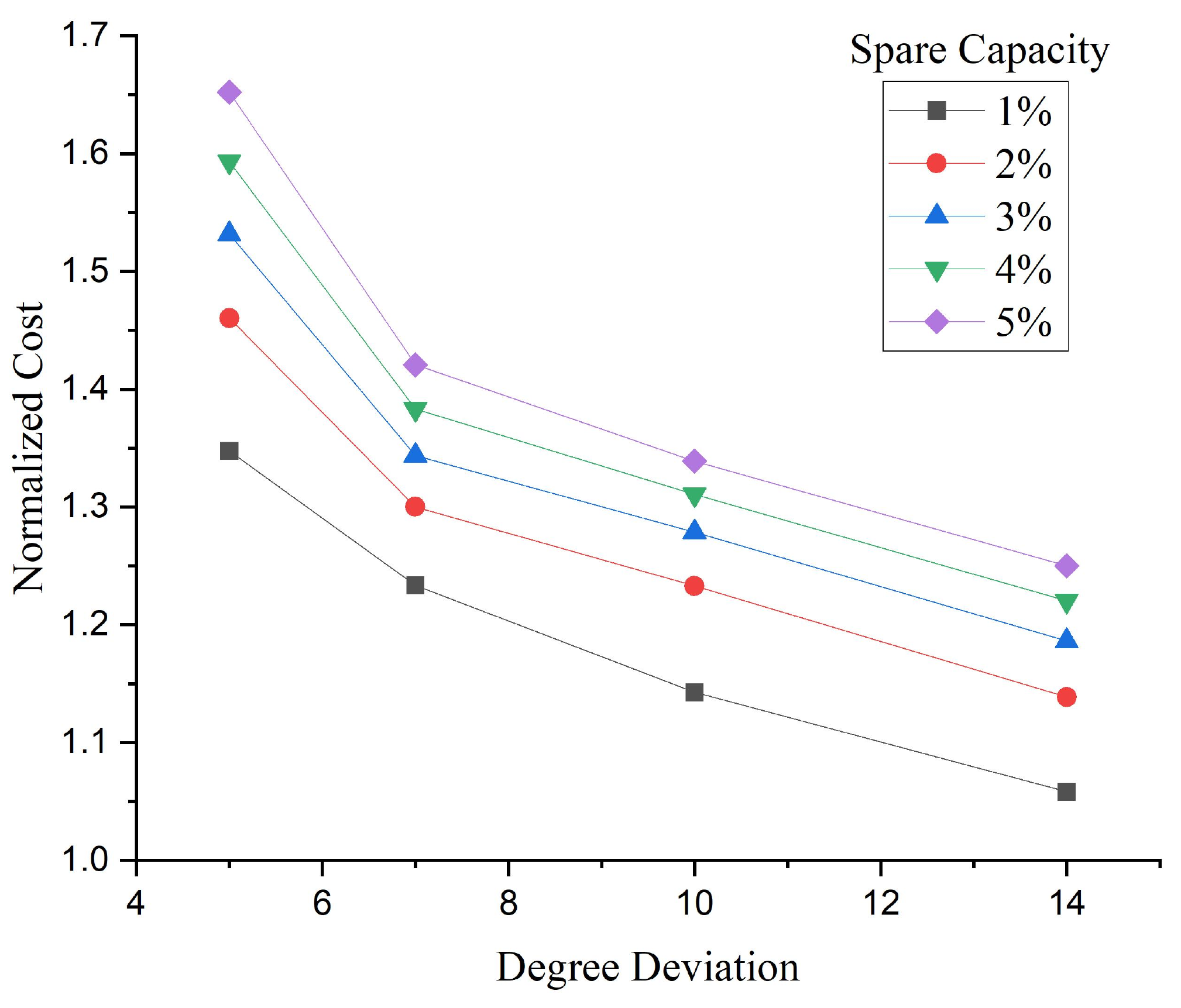}
\caption{The normalized cost of spare capacity as a function of degree deviation is shown in the figure for USAir97 when betweenness centrality was used as the method of node disruption. Normalized cost of spare capacity of 1 (closed squares), 2 (closed circles), 3 (closed triangles), 4 (closed inverted triangles) and 5 (closed diamonds) percent. Similar data for this and other networks is given in Supplementary Material.}
\label{fig:normalised-cost}
\end{figure}

The cost of spare capacity is a function of the degree deviation of the nodes in the network;  as we constrain the number of spare edges a node can accommodate, the algorithm chooses node pairs that result in an increasing sum of finite path lengths for the network, thereby increasing the cost of spare capacity. To better understand the implications of this cost, we define a metric termed as the \emph{normalized cost}, which is the cost of spare capacity at a given degree deviation divided (normalized) by that when there is no degree deviation imposed on the nodes in the network. From Figure~\ref{fig:normalised-cost}, it is evident that as we constrain the degree deviation of a network, there is a consequent increase in the normalised cost. Moreover, over a threshold of degree deviation, this increase in cost is non-linear and most likely prohibitive---as the degree deviation is constrained, there comes a point when it becomes infeasible to add further spare capacity. 

As an example, if the applied degree deviation constraint is two and all the nodes in either or both of the connected components have a degree of two, then it is not possible to add further spare capacity in those nodes and our algorithm returns a message to that effect. This was the case when the degree deviation was constrained at levels below that shown in Figure~\ref{fig:normalised-cost} for USAir97. Hence, it is necessary for the architects designing spare capacity to analyse the impact of degree deviation on normalized cost and come to an optimal solution that is specific for a given network.

\section{Discussion}
\label{sec:discussion}

In this study, we study the vulnerability of networks to targeted attacks, and propose a novel approach to mitigate the damage through a judicious re-wiring of the existing nodes. Using the examples of two important infrastructure networks, other real-world networks and canonical power-law networks, we demonstrate that these networks can be made significantly more robust, by exploiting the notion of \emph{spare capacity}.

Our algorithm is highly effective in increasing the robustness of networks; from all the real-world networks investigated in our study, the highest increase in robustness value at 5\% spare capacity was 1,593\% and 712\% when betweenness centrality and degree centrality were used to disrupt the networks respectively (mean value for all the analysed networks---449\% and 148\% respectively). To our knowledge, these numbers are significantly higher than that reported in the literature ~\citep{Schneider2011Mitigation,Li2019Maximizing,Louzada2013Smart,rong2018heuristic,jiang2011enhancing}.

When compared to the degree invariant approach of ES strategies in general and the onion network strategy in particular~\citep{Schneider2011Mitigation}, though one may argue that the edges are swapped and not added, one has to understand that swapping an edge (for e.g. either a power line or a fiber optic cable) between two pairs of nodes will have sizable cost implications. Moreover, the onion network suffers from another shortcoming---though the core of high degree nodes are relatively insulated from attack, the periphery of lower degree nodes have an increased attack vulnerability~\citep{Zeng2012Enhancing}. Our approach mitigates this limitation by taking a holistic view of the network. The spare capacity deployed in the network is also flexible whereby it can be selectively activated at specific regions of the network depending on the nature of node/nodes disrupted.

What does spare capacity in these networks imply? In the case of USAir97, it informs us the way to efficiency increase the capacity of a given number of airports (nodes) to a given extent (degree deviation) so that, in the event of an airport hub being disrupted due to bad weather or a terror attack, the air infrastructure of a region can function relatively smoothly till the disrupted airport is rendered operational. Moreover, at 5\% spare capacity, USAir97 more or less retains its original functionality (Figure~\ref{fig:robustness-us-air})--- a tremendous operational advantage for a real world air network. In the case of 1138-Bus, the implication is similar but of a different commodity---power. In other words, our algorithm distributes the traffic of the high centrality nodes, better termed as hubs, through various other nodes in the network \textit{i.e.} it \emph{randomizes} the hubs.

\subsection*{Biological Networks are a cornucopia of spare capacity}
\label{sub:biological-networks-are-a-cornucopia-of-spare-capacity}

One of the most well studied biological spare capacity is that of the bow tie architecture~\citep{Whitacre2012Biological} in which various substrates are connected to a core of conserved pathways. In this architecture, the loss of one substrate (glucose) does not necessarily imply the loss of informational integrity of the network (life). Rather, a spare capacity (lactose) kicks in, which sustains the life of the organism.

In a thought-provoking work~\citep{Wagner2005Distributed}, Wagner portrays the concept of distributed robustness as a different aspect of biological robustness as compared to redundancy. Unlike the latter, distributed robustness is a phenomenon in which \emph{distributed and different} parts of a biological network play a compensatory role for a given loss of function, akin to the spare capacity of a network. For example, the loss of Glucose-6-phosphate (G6P) dehydrogenase gene, whose product is an important enzyme in the oxidative pentose pathway shunt, results in the growth of the organism at near wild type levels~\citep{friesen1988escherichia}. However, there is a significant shift in the metabolite network of the organism, in which the loss of NADPH due to G6P mutation is compensated for by an increased flux of the metabolite through the tricarboxylic acid cycle.

The phenomenon of distributed robustness is also evident in the weak gene perturbation effects seen in biological systems in which the disruption of a gene does not significantly affect the phenotype of an organism~\citep{gu2003role,conant2004duplicate,giaever2002functional}. In the aforementioned work~\citep{Wagner2005Distributed}, it was noted that the cause of this effect cannot be attributed to gene duplication alone and some form of distributed robustness, where a gene product compensates for the loss of function of a different gene product, is likely.

One of the most destructive effects of spare capacity is evident in cancer. In this disease, in addition to the usual drug resistance mechanisms of drug efflux, target reactivation and drug modification and neutralization that are also the hallmarks of multi-drug resistant bacteria~\citep{Petchiappan2017Antibiotic}, distributed robustness also plays an important part in therapy resistance and disease progression. Adaptive kinome response, in which inhibition of a kinase to target a specific cell signaling pathway in cancer results in the activation of other kinases that compensate for the loss of function of the targeted kinase, is a typical example of spare capacity~\citep{Johnson2014Molecular}. Another example of this phenomenon is the MEK-ERK1/2 pathway that is activated in colorectal cancer. Targeting this pathway results in the activation of ERK 5 pathway that compensates for the role of ERK1/2 pathway. An interesting observation here is that ERK1/2 inhibits ERK 5 through negative feedback control---in other words, it keeps the spare capacity dormant~\citep{deJong2016ERK5}.
\\
\\
\subsection*{Future Perspectives}\label{future-perspectives}

The modus operandi of connecting the two largest connected components in a network with the lowest possible cost represents the core of our algorithm, the extension of which to biological networks leads to some interesting observations. Biological networks are different from infrastructure networks in one crucial aspect---the edges between two nodes in a biological network are relatively much more flexible and therefore more dynamic. While it may not be possible for a power line to dynamically reorient itself between different stations, a covalent or a non-covalent interaction between two nodes in a biological network can do so with minimal cost. In other words, the already available edges can also reorient themselves at a minimal cost in combination with \textit{de novo} addition of new edges. Therefore, in the event of a node disruption, be it in the form of a gene deletion, non-functionality of a transcription factor~\citep{Wu2015Functional} or simply the presence of two connected components that need to find the minimal cost of association between the two in, say, protein folding, the use of our algorithm in these scenarios may lead to some interesting results. 

In closing, we see two main contributions of our work. First, we present an algorithm to identify the most optimal utilisation of spare capacity, establishing new connections between existing nodes, in order to achieve a significant increase in robustness. Secondly, as a measure of the efficiency of the added spare capacity, we have introduced the concept of normalized cost for the addition of spare capacity between different node pairs in a complex network. Ultimately, the concept of spare capacity, borrowed from nature, can be a versatile tool to understand and manipulate networks, both biological and non-biological.

\bibliography{apssamp}

\begin{thebibliography}{28}%
\makeatletter
\providecommand \@ifxundefined [1]{%
 \@ifx{#1\undefined}
}%
\providecommand \@ifnum [1]{%
 \ifnum #1\expandafter \@firstoftwo
 \else \expandafter \@secondoftwo
 \fi
}%
\providecommand \@ifx [1]{%
 \ifx #1\expandafter \@firstoftwo
 \else \expandafter \@secondoftwo
 \fi
}%
\providecommand \natexlab [1]{#1}%
\providecommand \enquote  [1]{``#1''}%
\providecommand \bibnamefont  [1]{#1}%
\providecommand \bibfnamefont [1]{#1}%
\providecommand \citenamefont [1]{#1}%
\providecommand \href@noop [0]{\@secondoftwo}%
\providecommand \href [0]{\begingroup \@sanitize@url \@href}%
\providecommand \@href[1]{\@@startlink{#1}\@@href}%
\providecommand \@@href[1]{\endgroup#1\@@endlink}%
\providecommand \@sanitize@url [0]{\catcode `\\12\catcode `\$12\catcode
  `\&12\catcode `\#12\catcode `\^12\catcode `\_12\catcode `\%12\relax}%
\providecommand \@@startlink[1]{}%
\providecommand \@@endlink[0]{}%
\providecommand \url  [0]{\begingroup\@sanitize@url \@url }%
\providecommand \@url [1]{\endgroup\@href {#1}{\urlprefix }}%
\providecommand \urlprefix  [0]{URL }%
\providecommand \Eprint [0]{\href }%
\providecommand \doibase [0]{https://doi.org/}%
\providecommand \selectlanguage [0]{\@gobble}%
\providecommand \bibinfo  [0]{\@secondoftwo}%
\providecommand \bibfield  [0]{\@secondoftwo}%
\providecommand \translation [1]{[#1]}%
\providecommand \BibitemOpen [0]{}%
\providecommand \bibitemStop [0]{}%
\providecommand \bibitemNoStop [0]{.\EOS\space}%
\providecommand \EOS [0]{\spacefactor3000\relax}%
\providecommand \BibitemShut  [1]{\csname bibitem#1\endcsname}%
\let\auto@bib@innerbib\@empty
\bibitem [{\citenamefont {Albert}\ \emph {et~al.}(2000)\citenamefont {Albert},
  \citenamefont {Jeong},\ and\ \citenamefont {Barabasi}}]{Albert2000Error}%
  \BibitemOpen
  \bibfield  {author} {\bibinfo {author} {\bibfnamefont {R.}~\bibnamefont
  {Albert}}, \bibinfo {author} {\bibfnamefont {H.}~\bibnamefont {Jeong}},\ and\
  \bibinfo {author} {\bibfnamefont {A.-L.}\ \bibnamefont {Barabasi}},\
  }\bibfield  {title} {\bibinfo {title} {Error and attack tolerance of complex
  networks},\ }\href {https://doi.org/10.1038/35019019} {\bibfield  {journal}
  {\bibinfo  {journal} {Nature}\ }\textbf {\bibinfo {volume} {406}},\ \bibinfo
  {pages} {378} (\bibinfo {year} {2000})},\ \Eprint
  {https://arxiv.org/abs/cond-mat/0008064} {arXiv:cond-mat/0008064}
  \BibitemShut {NoStop}%
\bibitem [{\citenamefont {Barab{\'a}si}\ and\ \citenamefont
  {Albert}(1999)}]{Barabasi1999Emergence}%
  \BibitemOpen
  \bibfield  {author} {\bibinfo {author} {\bibfnamefont {A.-L.}\ \bibnamefont
  {Barab{\'a}si}}\ and\ \bibinfo {author} {\bibfnamefont {R.}~\bibnamefont
  {Albert}},\ }\bibfield  {title} {\bibinfo {title} {Emergence of scaling in
  random networks},\ }\href {https://doi.org/10.1126/science.286.5439.509}
  {\bibfield  {journal} {\bibinfo  {journal} {Science}\ }\textbf {\bibinfo
  {volume} {286}},\ \bibinfo {pages} {509} (\bibinfo {year} {1999})},\ \Eprint
  {https://arxiv.org/abs/cond-mat/9910332} {arXiv:cond-mat/9910332}
  \BibitemShut {NoStop}%
\bibitem [{\citenamefont {Caldarelli}(2007)}]{caldarelli2007scale}%
  \BibitemOpen
  \bibfield  {author} {\bibinfo {author} {\bibfnamefont {G.}~\bibnamefont
  {Caldarelli}},\ }\href@noop {} {\emph {\bibinfo {title} {Scale-free networks:
  complex webs in nature and technology}}}\ (\bibinfo  {publisher} {Oxford
  University Press},\ \bibinfo {year} {2007})\BibitemShut {NoStop}%
\bibitem [{\citenamefont {Cohen}\ \emph {et~al.}(2001)\citenamefont {Cohen},
  \citenamefont {Erez}, \citenamefont {Ben-Avraham},\ and\ \citenamefont
  {Havlin}}]{cohen2001breakdown}%
  \BibitemOpen
  \bibfield  {author} {\bibinfo {author} {\bibfnamefont {R.}~\bibnamefont
  {Cohen}}, \bibinfo {author} {\bibfnamefont {K.}~\bibnamefont {Erez}},
  \bibinfo {author} {\bibfnamefont {D.}~\bibnamefont {Ben-Avraham}},\ and\
  \bibinfo {author} {\bibfnamefont {S.}~\bibnamefont {Havlin}},\ }\bibfield
  {title} {\bibinfo {title} {Breakdown of the internet under intentional
  attack},\ }\href@noop {} {\bibfield  {journal} {\bibinfo  {journal} {Physical
  review letters}\ }\textbf {\bibinfo {volume} {86}},\ \bibinfo {pages} {3682}
  (\bibinfo {year} {2001})}\BibitemShut {NoStop}%
\bibitem [{\citenamefont {Albert}\ \emph {et~al.}(2004)\citenamefont {Albert},
  \citenamefont {Albert},\ and\ \citenamefont
  {Nakarado}}]{albert2004structural}%
  \BibitemOpen
  \bibfield  {author} {\bibinfo {author} {\bibfnamefont {R.}~\bibnamefont
  {Albert}}, \bibinfo {author} {\bibfnamefont {I.}~\bibnamefont {Albert}},\
  and\ \bibinfo {author} {\bibfnamefont {G.~L.}\ \bibnamefont {Nakarado}},\
  }\bibfield  {title} {\bibinfo {title} {Structural vulnerability of the north
  american power grid},\ }\href@noop {} {\bibfield  {journal} {\bibinfo
  {journal} {Physical review E}\ }\textbf {\bibinfo {volume} {69}},\ \bibinfo
  {pages} {025103} (\bibinfo {year} {2004})}\BibitemShut {NoStop}%
\bibitem [{\citenamefont {Schneider}\ \emph {et~al.}(2011)\citenamefont
  {Schneider}, \citenamefont {Moreira}, \citenamefont {Andrade}, \citenamefont
  {Havlin},\ and\ \citenamefont {Herrmann}}]{Schneider2011Mitigation}%
  \BibitemOpen
  \bibfield  {author} {\bibinfo {author} {\bibfnamefont {C.~M.}\ \bibnamefont
  {Schneider}}, \bibinfo {author} {\bibfnamefont {A.~A.}\ \bibnamefont
  {Moreira}}, \bibinfo {author} {\bibfnamefont {J.~S.}\ \bibnamefont
  {Andrade}}, \bibinfo {author} {\bibfnamefont {S.}~\bibnamefont {Havlin}},\
  and\ \bibinfo {author} {\bibfnamefont {H.~J.}\ \bibnamefont {Herrmann}},\
  }\bibfield  {title} {\bibinfo {title} {Mitigation of malicious attacks on
  networks},\ }\href {https://doi.org/10.1073/pnas.1009440108} {\bibfield
  {journal} {\bibinfo  {journal} {Proceedings of the National Academy of
  Sciences}\ }\textbf {\bibinfo {volume} {108}},\ \bibinfo {pages} {3838}
  (\bibinfo {year} {2011})}\BibitemShut {NoStop}%
\bibitem [{\citenamefont {Louzada}\ \emph {et~al.}(2013)\citenamefont
  {Louzada}, \citenamefont {Daolio}, \citenamefont {Herrmann},\ and\
  \citenamefont {Tomassini}}]{Louzada2013Smart}%
  \BibitemOpen
  \bibfield  {author} {\bibinfo {author} {\bibfnamefont {V.~H.~P.}\
  \bibnamefont {Louzada}}, \bibinfo {author} {\bibfnamefont {F.}~\bibnamefont
  {Daolio}}, \bibinfo {author} {\bibfnamefont {H.~J.}\ \bibnamefont
  {Herrmann}},\ and\ \bibinfo {author} {\bibfnamefont {M.}~\bibnamefont
  {Tomassini}},\ }\bibfield  {title} {\bibinfo {title} {Smart rewiring for
  network robustness},\ }\href {https://doi.org/10.1093/comnet/cnt010}
  {\bibfield  {journal} {\bibinfo  {journal} {Journal of Complex Networks}\
  }\textbf {\bibinfo {volume} {1}},\ \bibinfo {pages} {150} (\bibinfo {year}
  {2013})}\BibitemShut {NoStop}%
\bibitem [{\citenamefont {Kr{\'o}l}\ \emph {et~al.}(2015)\citenamefont
  {Kr{\'o}l}, \citenamefont {Fay},\ and\ \citenamefont
  {Gabry{\'s}}}]{Krol2015Propagation}%
  \BibitemOpen
  \bibinfo {editor} {\bibfnamefont {D.}~\bibnamefont {Kr{\'o}l}}, \bibinfo
  {editor} {\bibfnamefont {D.}~\bibnamefont {Fay}},\ and\ \bibinfo {editor}
  {\bibfnamefont {B.}~\bibnamefont {Gabry{\'s}}},\ eds.,\ \href
  {https://doi.org/10.1007/978-3-319-15916-4} {\emph {\bibinfo {title}
  {Propagation {{Phenomena}} in {{Real World Networks}}}}},\ Intelligent
  {{Systems Reference Library}}\ (\bibinfo  {publisher} {{Springer
  International Publishing}},\ \bibinfo {year} {2015})\BibitemShut {NoStop}%
\bibitem [{\citenamefont {Rong}\ and\ \citenamefont
  {Liu}(2018)}]{rong2018heuristic}%
  \BibitemOpen
  \bibfield  {author} {\bibinfo {author} {\bibfnamefont {L.}~\bibnamefont
  {Rong}}\ and\ \bibinfo {author} {\bibfnamefont {J.}~\bibnamefont {Liu}},\
  }\bibfield  {title} {\bibinfo {title} {A heuristic algorithm for enhancing
  the robustness of scale-free networks based on edge classification},\
  }\href@noop {} {\bibfield  {journal} {\bibinfo  {journal} {Physica A:
  Statistical Mechanics and its Applications}\ }\textbf {\bibinfo {volume}
  {503}},\ \bibinfo {pages} {503} (\bibinfo {year} {2018})}\BibitemShut
  {NoStop}%
\bibitem [{\citenamefont {Wu}\ and\ \citenamefont {Holme}(2011)}]{wu2011onion}%
  \BibitemOpen
  \bibfield  {author} {\bibinfo {author} {\bibfnamefont {Z.-X.}\ \bibnamefont
  {Wu}}\ and\ \bibinfo {author} {\bibfnamefont {P.}~\bibnamefont {Holme}},\
  }\bibfield  {title} {\bibinfo {title} {Onion structure and network
  robustness},\ }\href@noop {} {\bibfield  {journal} {\bibinfo  {journal}
  {Physical Review E}\ }\textbf {\bibinfo {volume} {84}},\ \bibinfo {pages}
  {026106} (\bibinfo {year} {2011})}\BibitemShut {NoStop}%
\bibitem [{\citenamefont {Tanizawa}\ \emph {et~al.}(2012)\citenamefont
  {Tanizawa}, \citenamefont {Havlin},\ and\ \citenamefont
  {Stanley}}]{tanizawa2012robustness}%
  \BibitemOpen
  \bibfield  {author} {\bibinfo {author} {\bibfnamefont {T.}~\bibnamefont
  {Tanizawa}}, \bibinfo {author} {\bibfnamefont {S.}~\bibnamefont {Havlin}},\
  and\ \bibinfo {author} {\bibfnamefont {H.~E.}\ \bibnamefont {Stanley}},\
  }\bibfield  {title} {\bibinfo {title} {Robustness of onionlike correlated
  networks against targeted attacks},\ }\href@noop {} {\bibfield  {journal}
  {\bibinfo  {journal} {Physical Review E}\ }\textbf {\bibinfo {volume} {85}},\
  \bibinfo {pages} {046109} (\bibinfo {year} {2012})}\BibitemShut {NoStop}%
\bibitem [{\citenamefont {Li}\ \emph {et~al.}(2019)\citenamefont {Li},
  \citenamefont {Li}, \citenamefont {Tan}, \citenamefont {Cao}, \citenamefont
  {Chen}, \citenamefont {Cai}, \citenamefont {Lee},\ and\ \citenamefont
  {Pecht}}]{Li2019Maximizing}%
  \BibitemOpen
  \bibfield  {author} {\bibinfo {author} {\bibfnamefont {W.}~\bibnamefont
  {Li}}, \bibinfo {author} {\bibfnamefont {Y.}~\bibnamefont {Li}}, \bibinfo
  {author} {\bibfnamefont {Y.}~\bibnamefont {Tan}}, \bibinfo {author}
  {\bibfnamefont {Y.}~\bibnamefont {Cao}}, \bibinfo {author} {\bibfnamefont
  {C.}~\bibnamefont {Chen}}, \bibinfo {author} {\bibfnamefont {Y.}~\bibnamefont
  {Cai}}, \bibinfo {author} {\bibfnamefont {K.~Y.}\ \bibnamefont {Lee}},\ and\
  \bibinfo {author} {\bibfnamefont {M.}~\bibnamefont {Pecht}},\ }\bibfield
  {title} {\bibinfo {title} {Maximizing {{Network Resilience}} against
  {{Malicious Attacks}}},\ }\href {https://doi.org/10.1038/s41598-019-38781-7}
  {\bibfield  {journal} {\bibinfo  {journal} {Scientific Reports}\ }\textbf
  {\bibinfo {volume} {9}},\ \bibinfo {pages} {2261} (\bibinfo {year}
  {2019})}\BibitemShut {NoStop}%
\bibitem [{\citenamefont {Jiang}\ \emph {et~al.}(2011)\citenamefont {Jiang},
  \citenamefont {Liang},\ and\ \citenamefont {Guo}}]{jiang2011enhancing}%
  \BibitemOpen
  \bibfield  {author} {\bibinfo {author} {\bibfnamefont {Z.}~\bibnamefont
  {Jiang}}, \bibinfo {author} {\bibfnamefont {M.}~\bibnamefont {Liang}},\ and\
  \bibinfo {author} {\bibfnamefont {D.}~\bibnamefont {Guo}},\ }\bibfield
  {title} {\bibinfo {title} {Enhancing network performance by edge addition},\
  }\href@noop {} {\bibfield  {journal} {\bibinfo  {journal} {International
  Journal of Modern Physics C}\ }\textbf {\bibinfo {volume} {22}},\ \bibinfo
  {pages} {1211} (\bibinfo {year} {2011})}\BibitemShut {NoStop}%
\bibitem [{\citenamefont {Paul}\ \emph {et~al.}(2004)\citenamefont {Paul},
  \citenamefont {Tanizawa}, \citenamefont {Havlin},\ and\ \citenamefont
  {Stanley}}]{paul2004optimization}%
  \BibitemOpen
  \bibfield  {author} {\bibinfo {author} {\bibfnamefont {G.}~\bibnamefont
  {Paul}}, \bibinfo {author} {\bibfnamefont {T.}~\bibnamefont {Tanizawa}},
  \bibinfo {author} {\bibfnamefont {S.}~\bibnamefont {Havlin}},\ and\ \bibinfo
  {author} {\bibfnamefont {H.~E.}\ \bibnamefont {Stanley}},\ }\bibfield
  {title} {\bibinfo {title} {Optimization of robustness of complex networks},\
  }\href@noop {} {\bibfield  {journal} {\bibinfo  {journal} {The European
  Physical Journal B}\ }\textbf {\bibinfo {volume} {38}},\ \bibinfo {pages}
  {187} (\bibinfo {year} {2004})}\BibitemShut {NoStop}%
\bibitem [{\citenamefont {Zhao}\ and\ \citenamefont
  {Xu}(2009)}]{zhao2009enhancing}%
  \BibitemOpen
  \bibfield  {author} {\bibinfo {author} {\bibfnamefont {J.}~\bibnamefont
  {Zhao}}\ and\ \bibinfo {author} {\bibfnamefont {K.}~\bibnamefont {Xu}},\
  }\bibfield  {title} {\bibinfo {title} {Enhancing the robustness of scale-free
  networks},\ }\href@noop {} {\bibfield  {journal} {\bibinfo  {journal}
  {Journal of Physics A: Mathematical and Theoretical}\ }\textbf {\bibinfo
  {volume} {42}},\ \bibinfo {pages} {195003} (\bibinfo {year}
  {2009})}\BibitemShut {NoStop}%
\bibitem [{\citenamefont {Rossi}\ and\ \citenamefont {Ahmed}(2015)}]{nr}%
  \BibitemOpen
  \bibfield  {author} {\bibinfo {author} {\bibfnamefont {R.~A.}\ \bibnamefont
  {Rossi}}\ and\ \bibinfo {author} {\bibfnamefont {N.~K.}\ \bibnamefont
  {Ahmed}},\ }\bibfield  {title} {\bibinfo {title} {The network data repository
  with interactive graph analytics and visualization},\ }in\ \href
  {https://networkrepository.com} {\emph {\bibinfo {booktitle} {AAAI}}}\
  (\bibinfo {year} {2015})\BibitemShut {NoStop}%
\bibitem [{\citenamefont {Holme}\ \emph {et~al.}(2002)\citenamefont {Holme},
  \citenamefont {Kim}, \citenamefont {Yoon},\ and\ \citenamefont
  {Han}}]{holme2002attack}%
  \BibitemOpen
  \bibfield  {author} {\bibinfo {author} {\bibfnamefont {P.}~\bibnamefont
  {Holme}}, \bibinfo {author} {\bibfnamefont {B.~J.}\ \bibnamefont {Kim}},
  \bibinfo {author} {\bibfnamefont {C.~N.}\ \bibnamefont {Yoon}},\ and\
  \bibinfo {author} {\bibfnamefont {S.~K.}\ \bibnamefont {Han}},\ }\bibfield
  {title} {\bibinfo {title} {Attack vulnerability of complex networks},\
  }\href@noop {} {\bibfield  {journal} {\bibinfo  {journal} {Physical review
  E}\ }\textbf {\bibinfo {volume} {65}},\ \bibinfo {pages} {056109} (\bibinfo
  {year} {2002})}\BibitemShut {NoStop}%
\bibitem [{\citenamefont {Zeng}\ and\ \citenamefont
  {Liu}(2012)}]{Zeng2012Enhancing}%
  \BibitemOpen
  \bibfield  {author} {\bibinfo {author} {\bibfnamefont {A.}~\bibnamefont
  {Zeng}}\ and\ \bibinfo {author} {\bibfnamefont {W.}~\bibnamefont {Liu}},\
  }\bibfield  {title} {\bibinfo {title} {Enhancing network robustness against
  malicious attacks},\ }\href {https://doi.org/10.1103/PhysRevE.85.066130}
  {\bibfield  {journal} {\bibinfo  {journal} {Physical Review E}\ }\textbf
  {\bibinfo {volume} {85}},\ \bibinfo {pages} {066130} (\bibinfo {year}
  {2012})}\BibitemShut {NoStop}%
\bibitem [{\citenamefont {Whitacre}(2012)}]{Whitacre2012Biological}%
  \BibitemOpen
  \bibfield  {author} {\bibinfo {author} {\bibfnamefont {J.~M.}\ \bibnamefont
  {Whitacre}},\ }\bibfield  {title} {\bibinfo {title} {Biological
  {{Robustness}}: {{Paradigms}}, {{Mechanisms}}, and {{Systems Principles}}},\
  }\bibfield  {journal} {\bibinfo  {journal} {Frontiers in Genetics}\ }\textbf
  {\bibinfo {volume} {0}},\ \href {https://doi.org/10.3389/fgene.2012.00067}
  {10.3389/fgene.2012.00067} (\bibinfo {year} {2012})\BibitemShut {NoStop}%
\bibitem [{\citenamefont {Wagner}(2005)}]{Wagner2005Distributed}%
  \BibitemOpen
  \bibfield  {author} {\bibinfo {author} {\bibfnamefont {A.}~\bibnamefont
  {Wagner}},\ }\bibfield  {title} {\bibinfo {title} {Distributed robustness
  versus redundancy as causes of mutational robustness.},\ }\href
  {https://doi.org/10.1002/bies.20170} {\bibfield  {journal} {\bibinfo
  {journal} {Bioessays}\ }\textbf {\bibinfo {volume} {27}},\ \bibinfo {pages}
  {176} (\bibinfo {year} {2005})}\BibitemShut {NoStop}%
\bibitem [{\citenamefont {Friesen}(1988)}]{friesen1988escherichia}%
  \BibitemOpen
  \bibfield  {author} {\bibinfo {author} {\bibfnamefont {J.~D.}\ \bibnamefont
  {Friesen}},\ }\bibfield  {title} {\bibinfo {title} {Escherichia coli and
  salmonella typhimurium: Cellular and molecular biology},\ }\href@noop {}
  {\bibfield  {journal} {\bibinfo  {journal} {Science}\ }\textbf {\bibinfo
  {volume} {240}},\ \bibinfo {pages} {1678} (\bibinfo {year}
  {1988})}\BibitemShut {NoStop}%
\bibitem [{\citenamefont {Gu}\ \emph {et~al.}(2003)\citenamefont {Gu},
  \citenamefont {Steinmetz}, \citenamefont {Gu}, \citenamefont {Scharfe},
  \citenamefont {Davis},\ and\ \citenamefont {Li}}]{gu2003role}%
  \BibitemOpen
  \bibfield  {author} {\bibinfo {author} {\bibfnamefont {Z.}~\bibnamefont
  {Gu}}, \bibinfo {author} {\bibfnamefont {L.~M.}\ \bibnamefont {Steinmetz}},
  \bibinfo {author} {\bibfnamefont {X.}~\bibnamefont {Gu}}, \bibinfo {author}
  {\bibfnamefont {C.}~\bibnamefont {Scharfe}}, \bibinfo {author} {\bibfnamefont
  {R.~W.}\ \bibnamefont {Davis}},\ and\ \bibinfo {author} {\bibfnamefont
  {W.-H.}\ \bibnamefont {Li}},\ }\bibfield  {title} {\bibinfo {title} {Role of
  duplicate genes in genetic robustness against null mutations},\ }\href@noop
  {} {\bibfield  {journal} {\bibinfo  {journal} {Nature}\ }\textbf {\bibinfo
  {volume} {421}},\ \bibinfo {pages} {63} (\bibinfo {year} {2003})}\BibitemShut
  {NoStop}%
\bibitem [{\citenamefont {Conant}\ and\ \citenamefont
  {Wagner}(2004)}]{conant2004duplicate}%
  \BibitemOpen
  \bibfield  {author} {\bibinfo {author} {\bibfnamefont {G.~C.}\ \bibnamefont
  {Conant}}\ and\ \bibinfo {author} {\bibfnamefont {A.}~\bibnamefont
  {Wagner}},\ }\bibfield  {title} {\bibinfo {title} {Duplicate genes and
  robustness to transient gene knock-downs in caenorhabditis elegans},\
  }\href@noop {} {\bibfield  {journal} {\bibinfo  {journal} {Proceedings of the
  Royal Society of London. Series B: Biological Sciences}\ }\textbf {\bibinfo
  {volume} {271}},\ \bibinfo {pages} {89} (\bibinfo {year} {2004})}\BibitemShut
  {NoStop}%
\bibitem [{\citenamefont {Giaever}\ \emph {et~al.}(2002)\citenamefont
  {Giaever}, \citenamefont {Chu}, \citenamefont {Ni}, \citenamefont {Connelly},
  \citenamefont {Riles}, \citenamefont {V{\'e}ronneau}, \citenamefont {Dow},
  \citenamefont {Lucau-Danila}, \citenamefont {Anderson}, \citenamefont
  {Andr{\'e}} \emph {et~al.}}]{giaever2002functional}%
  \BibitemOpen
  \bibfield  {author} {\bibinfo {author} {\bibfnamefont {G.}~\bibnamefont
  {Giaever}}, \bibinfo {author} {\bibfnamefont {A.~M.}\ \bibnamefont {Chu}},
  \bibinfo {author} {\bibfnamefont {L.}~\bibnamefont {Ni}}, \bibinfo {author}
  {\bibfnamefont {C.}~\bibnamefont {Connelly}}, \bibinfo {author}
  {\bibfnamefont {L.}~\bibnamefont {Riles}}, \bibinfo {author} {\bibfnamefont
  {S.}~\bibnamefont {V{\'e}ronneau}}, \bibinfo {author} {\bibfnamefont
  {S.}~\bibnamefont {Dow}}, \bibinfo {author} {\bibfnamefont {A.}~\bibnamefont
  {Lucau-Danila}}, \bibinfo {author} {\bibfnamefont {K.}~\bibnamefont
  {Anderson}}, \bibinfo {author} {\bibfnamefont {B.}~\bibnamefont {Andr{\'e}}},
  \emph {et~al.},\ }\bibfield  {title} {\bibinfo {title} {Functional profiling
  of the saccharomyces cerevisiae genome},\ }\href@noop {} {\bibfield
  {journal} {\bibinfo  {journal} {nature}\ }\textbf {\bibinfo {volume} {418}},\
  \bibinfo {pages} {387} (\bibinfo {year} {2002})}\BibitemShut {NoStop}%
\bibitem [{\citenamefont {Petchiappan}\ and\ \citenamefont
  {Chatterji}(2017)}]{Petchiappan2017Antibiotic}%
  \BibitemOpen
  \bibfield  {author} {\bibinfo {author} {\bibfnamefont {A.}~\bibnamefont
  {Petchiappan}}\ and\ \bibinfo {author} {\bibfnamefont {D.}~\bibnamefont
  {Chatterji}},\ }\bibfield  {title} {\bibinfo {title} {Antibiotic
  {{Resistance}}: {{Current Perspectives}}},\ }\href
  {https://doi.org/10.1021/acsomega.7b01368} {\bibfield  {journal} {\bibinfo
  {journal} {ACS Omega}\ }\textbf {\bibinfo {volume} {2}},\ \bibinfo {pages}
  {7400} (\bibinfo {year} {2017})}\BibitemShut {NoStop}%
\bibitem [{\citenamefont {Johnson}\ \emph {et~al.}(2014)\citenamefont
  {Johnson}, \citenamefont {Stuhlmiller}, \citenamefont {Angus}, \citenamefont
  {Zawistowski},\ and\ \citenamefont {Graves}}]{Johnson2014Molecular}%
  \BibitemOpen
  \bibfield  {author} {\bibinfo {author} {\bibfnamefont {G.~L.}\ \bibnamefont
  {Johnson}}, \bibinfo {author} {\bibfnamefont {T.~J.}\ \bibnamefont
  {Stuhlmiller}}, \bibinfo {author} {\bibfnamefont {S.~P.}\ \bibnamefont
  {Angus}}, \bibinfo {author} {\bibfnamefont {J.~S.}\ \bibnamefont
  {Zawistowski}},\ and\ \bibinfo {author} {\bibfnamefont {L.~M.}\ \bibnamefont
  {Graves}},\ }\bibfield  {title} {\bibinfo {title} {Molecular {{Pathways}}:
  {{Adaptive Kinome Reprogramming}} in {{Response}} to {{Targeted Inhibition}}
  of the {{BRAF}}-{{MEK}}-{{ERK Pathway}} in {{Cancer}}},\ }\href
  {https://doi.org/10.1158/1078-0432.CCR-13-1081} {\bibfield  {journal}
  {\bibinfo  {journal} {Clinical cancer research : an official journal of the
  American Association for Cancer Research}\ }\textbf {\bibinfo {volume}
  {20}},\ \bibinfo {pages} {2516} (\bibinfo {year} {2014})}\BibitemShut
  {NoStop}%
\bibitem [{\citenamefont {{de Jong}}\ \emph {et~al.}(2016)\citenamefont {{de
  Jong}}, \citenamefont {Taniguchi}, \citenamefont {Harris}, \citenamefont
  {Bertin}, \citenamefont {Takahashi}, \citenamefont {Duong}, \citenamefont
  {Campos}, \citenamefont {Powis}, \citenamefont {Corr}, \citenamefont
  {Karin},\ and\ \citenamefont {Raz}}]{deJong2016ERK5}%
  \BibitemOpen
  \bibfield  {author} {\bibinfo {author} {\bibfnamefont {P.~R.}\ \bibnamefont
  {{de Jong}}}, \bibinfo {author} {\bibfnamefont {K.}~\bibnamefont
  {Taniguchi}}, \bibinfo {author} {\bibfnamefont {A.~R.}\ \bibnamefont
  {Harris}}, \bibinfo {author} {\bibfnamefont {S.}~\bibnamefont {Bertin}},
  \bibinfo {author} {\bibfnamefont {N.}~\bibnamefont {Takahashi}}, \bibinfo
  {author} {\bibfnamefont {J.}~\bibnamefont {Duong}}, \bibinfo {author}
  {\bibfnamefont {A.~D.}\ \bibnamefont {Campos}}, \bibinfo {author}
  {\bibfnamefont {G.}~\bibnamefont {Powis}}, \bibinfo {author} {\bibfnamefont
  {M.}~\bibnamefont {Corr}}, \bibinfo {author} {\bibfnamefont {M.}~\bibnamefont
  {Karin}},\ and\ \bibinfo {author} {\bibfnamefont {E.}~\bibnamefont {Raz}},\
  }\bibfield  {title} {\bibinfo {title} {{{ERK5}} signalling rescues intestinal
  epithelial turnover and tumour cell proliferation upon {{ERK1}}/2
  abrogation},\ }\href {https://doi.org/10.1038/ncomms11551} {\bibfield
  {journal} {\bibinfo  {journal} {Nature Communications}\ }\textbf {\bibinfo
  {volume} {7}},\ \bibinfo {pages} {11551} (\bibinfo {year}
  {2016})}\BibitemShut {NoStop}%
\bibitem [{\citenamefont {Wu}\ and\ \citenamefont
  {Lai}(2015)}]{Wu2015Functional}%
  \BibitemOpen
  \bibfield  {author} {\bibinfo {author} {\bibfnamefont {W.-S.}\ \bibnamefont
  {Wu}}\ and\ \bibinfo {author} {\bibfnamefont {F.-J.}\ \bibnamefont {Lai}},\
  }\bibfield  {title} {\bibinfo {title} {Functional redundancy of transcription
  factors explains why most binding targets of a transcription factor are not
  affected when the transcription factor is knocked out},\ }\href
  {https://doi.org/10.1186/1752-0509-9-S6-S2} {\bibfield  {journal} {\bibinfo
  {journal} {BMC Systems Biology}\ }\textbf {\bibinfo {volume} {9}},\ \bibinfo
  {pages} {S2} (\bibinfo {year} {2015})}\BibitemShut {NoStop}%
\end{thebibliography}%

\clearpage
\onecolumngrid
\appendix
\section{Supplementary Material}

Table~S1 contains the list of networks that have been analysed in this study. Apart from infrastructure networks, a few other networks like collaboration networks, biological networks and crime networks have been analysed. Our comments on the biological networks in the main part of the article notwithstanding, our algorithm has utility in a variety of networks. For example, an analysis of an author collaboration network will tell an author/scientist the most efficient way to reach a given contact should a senior Professor in the network (a node with high centrality) retire. Similarly, as regards to a crime network, it will provide the law enforcement agencies a list of the secondary targets that are capable of reforming the network after the primary targets have been neutralised, so that they can act appropriately.

\setcounter{figure}{0}
\renewcommand{\thefigure}{S\arabic{figure}}
\section{Supplementary Figures}

\begin{figure}[h]
\centerline{\includegraphics[width=0.55\linewidth]{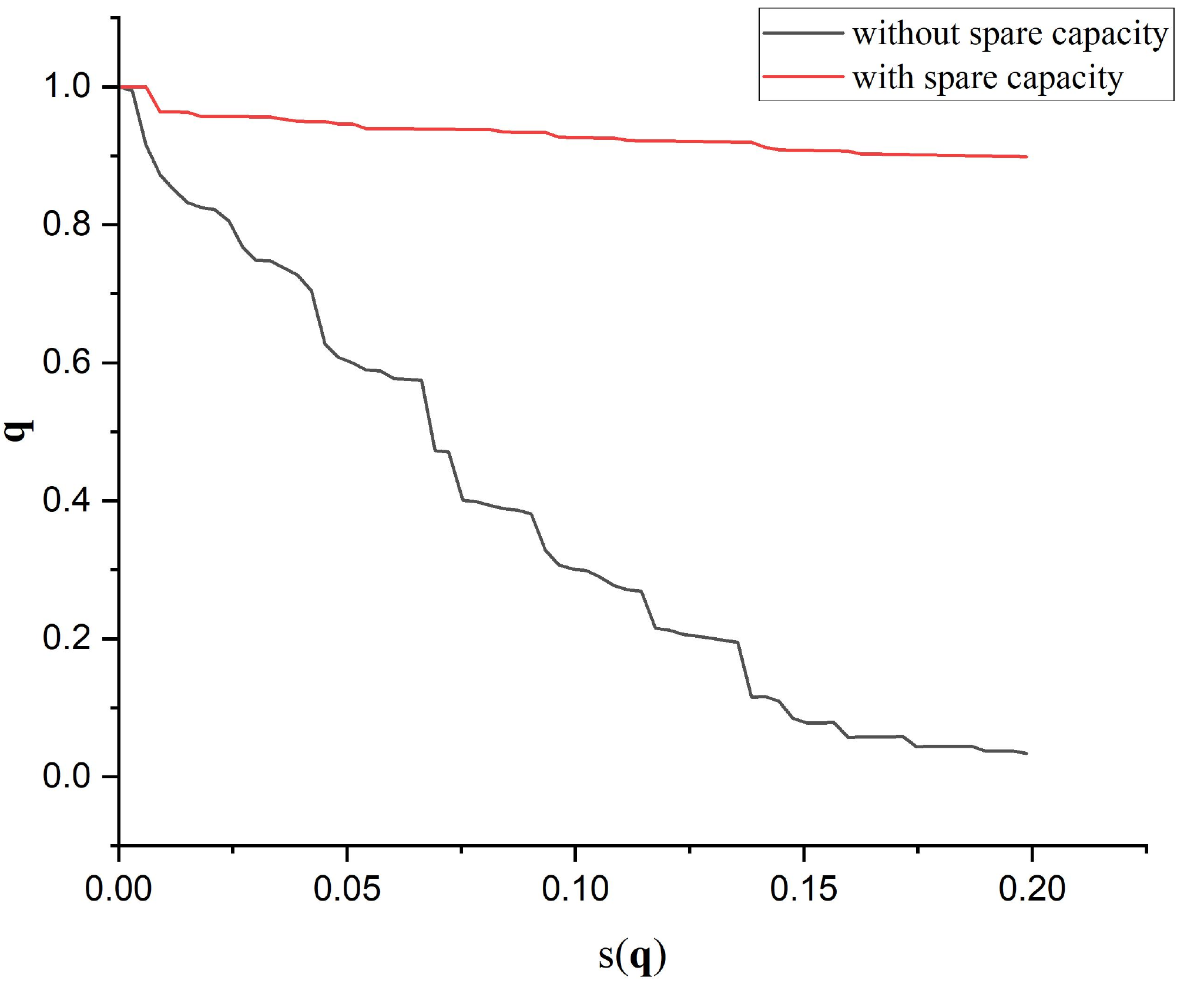}}
\caption{The robustness graph of USAir97 when betweenness
centrality was used as the mode of network disruption without and with 5\% Spare capacity (black and red lines respectively). Note the difference in the area under the curve between this figure
and Figure~\ref{fig:robustness-1138} in the main text. This figure leads to another important inference: at 5\% spare capacity, the USAir97 network will more or less retain its original functionality since the decrease in the robustness of the network is minimal (red plot). For a real world air network, this has profound implications. }
\label{fig:robustness-us-air}
\end{figure}

\begin{figure}
\centerline{\includegraphics[width=0.55\linewidth]{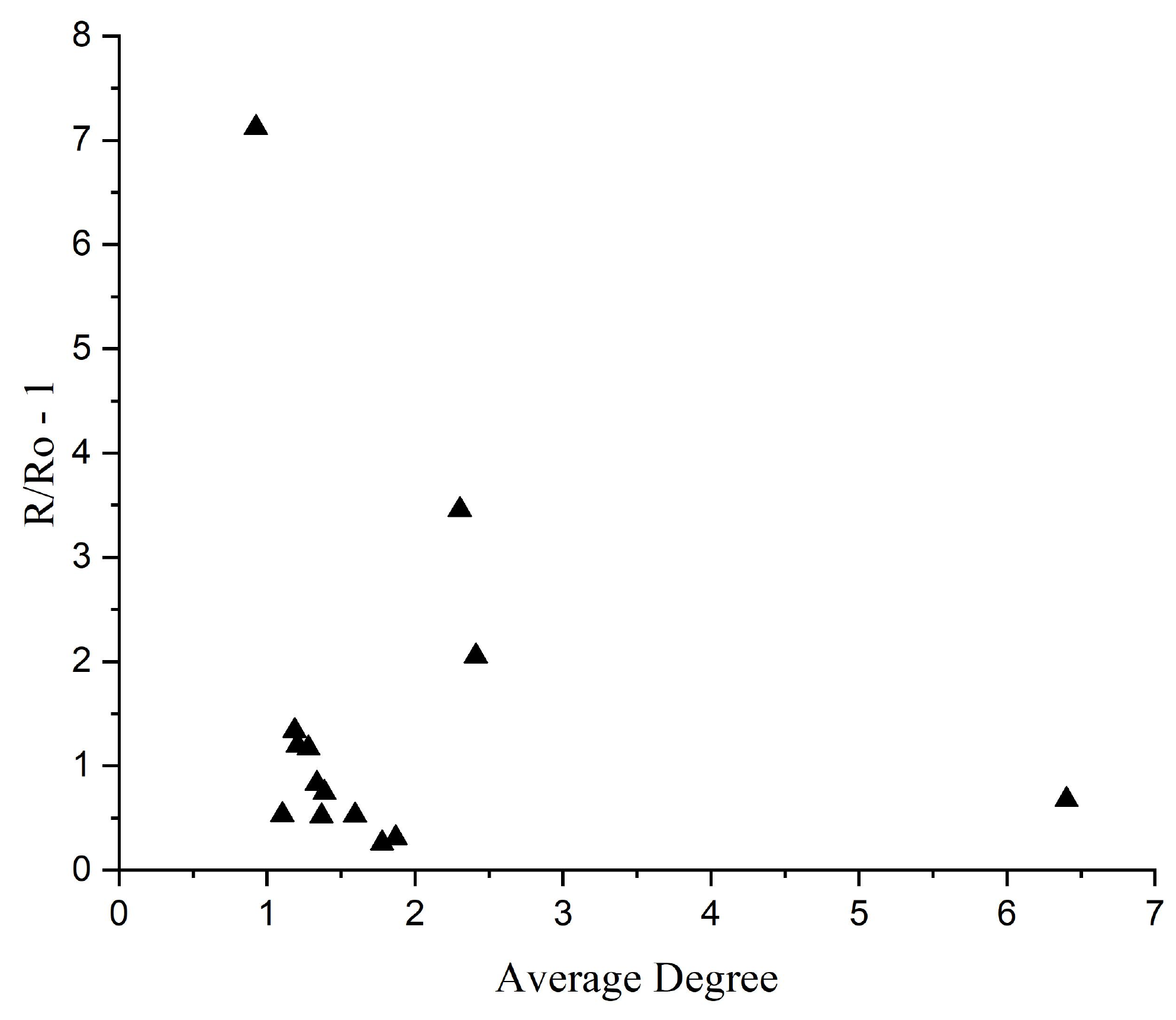}}

\caption{The robustness values of different real life networks analysed in this study, when degree centrality was used as the measure of node disruption. For more information on the networks refer to Supplementary Table S1.}
\label{fig:different-networks-dc}
\end{figure}

\begin{figure}[h]
\centerline{\includegraphics[width=0.55\linewidth]{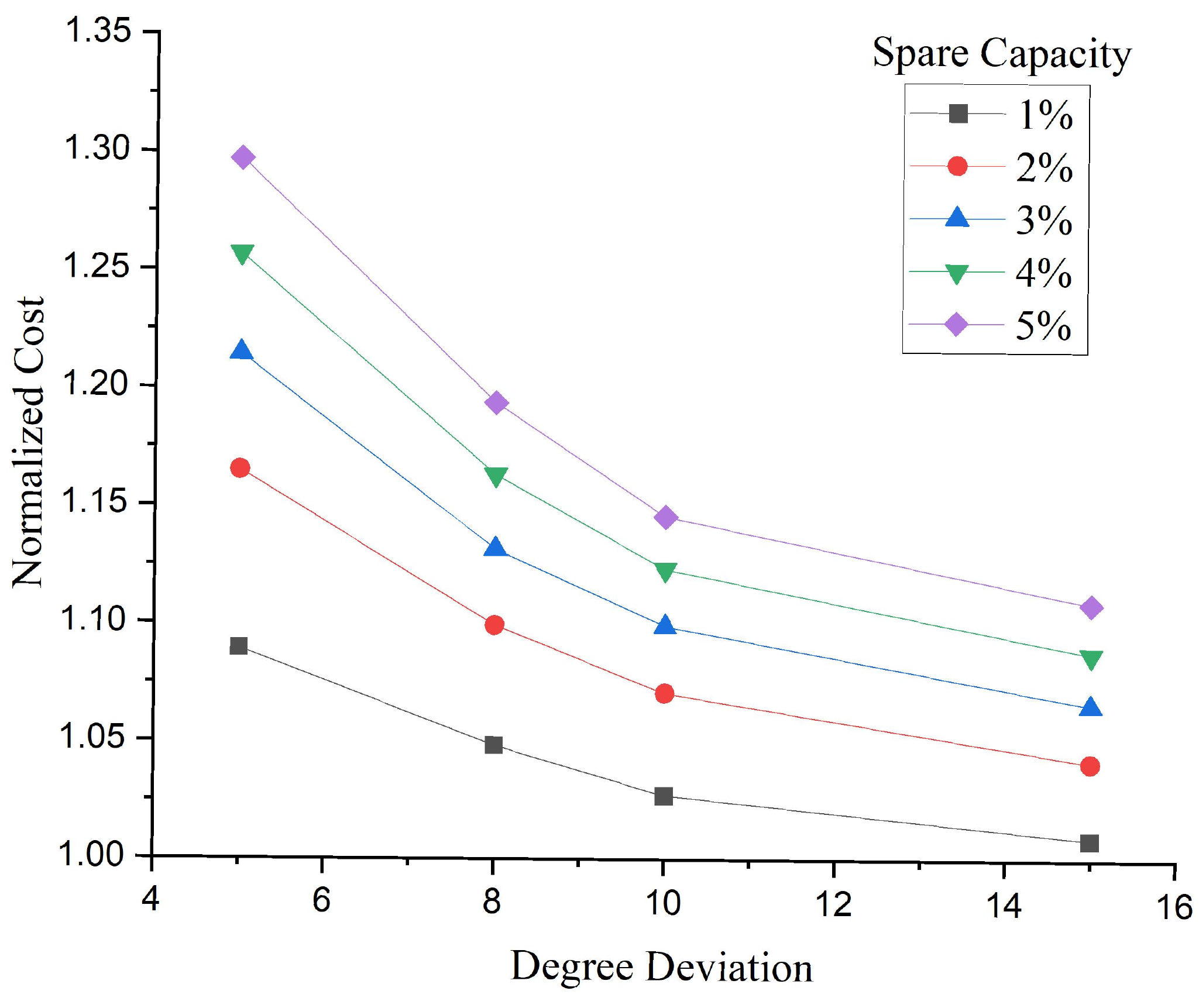}}

\caption{The normalized cost of spare capacity as a function of degree deviation is shown in the figure for USAir 97 when degree centrality was used as the method of node disruption. Normalised cost for spare capacity of 1 (closed squares), 2 (closed
circles), 3 (closed triangles), 4 (closed inverted triangles) and 5
(closed diamonds) percent.}
\label{fig:USAir97-DC-DD}
\end{figure}

\begin{figure}[h]
\centerline{\includegraphics[width=0.55\linewidth]{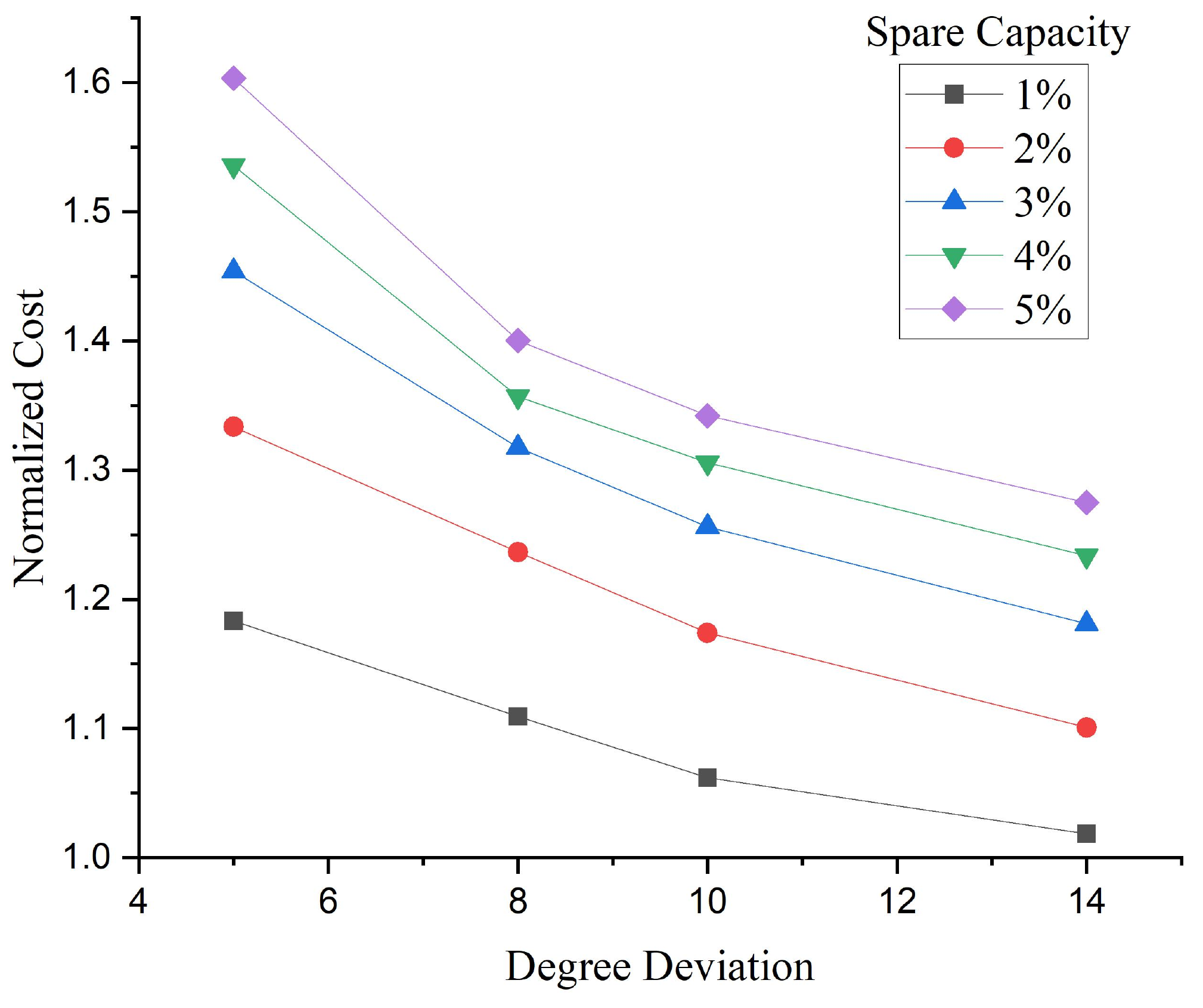}}

\caption{The normalized cost of spare capacity as a function of degree deviation is shown in the figure for 1138-Bus when betweenness centrality was used as the method of node disruption. Normalised cost for spare capacity of 1 (closed squares), 2 (closed
circles), 3 (closed triangles), 4 (closed inverted triangles) and 5
(closed diamonds) percent.}
\label{fig:1138-Bus-BC-DD}
\end{figure}

\begin{figure}[h]
\centerline{\includegraphics[width=0.55\linewidth]{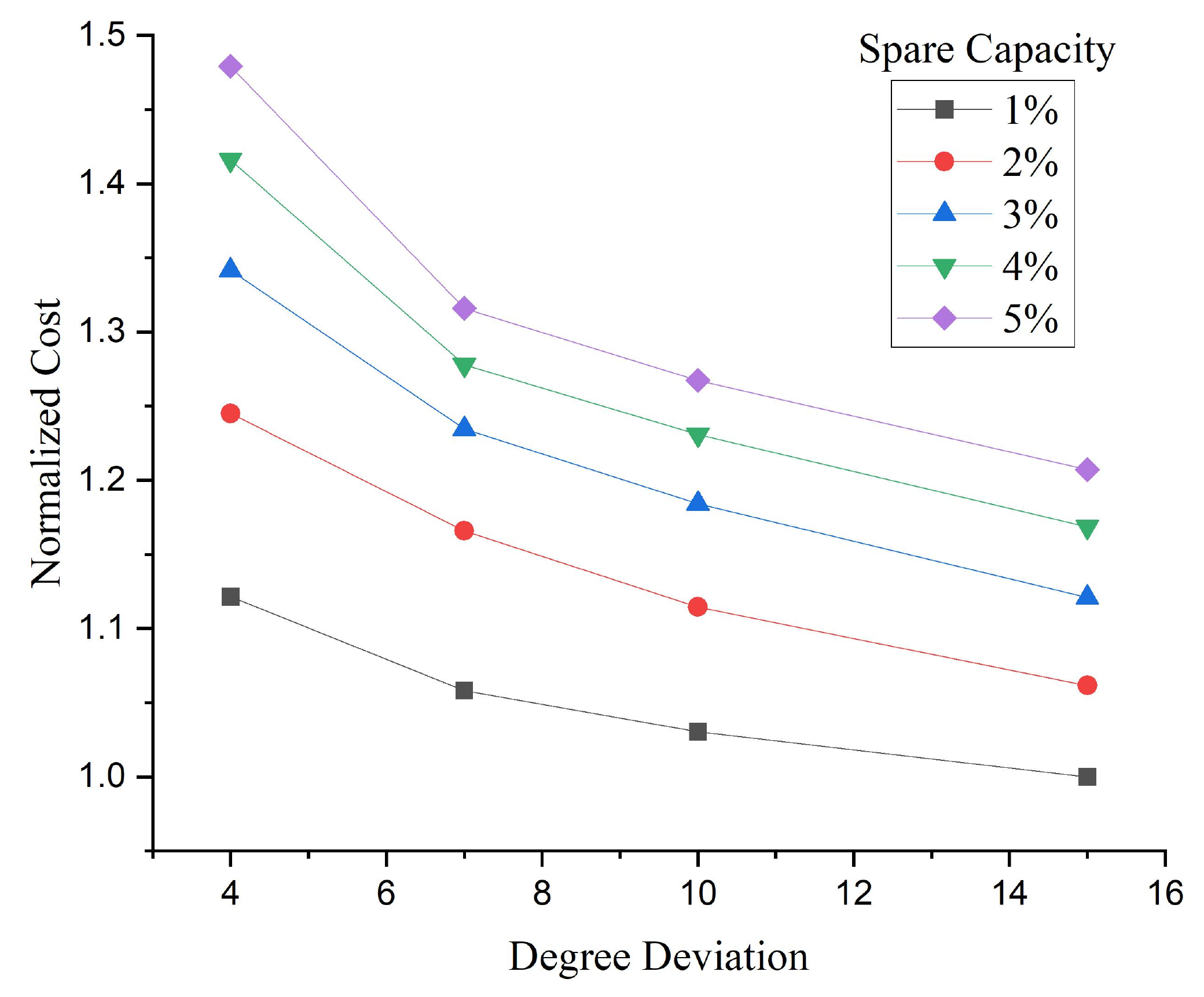}}

\caption{The normalized cost of spare capacity as a function of degree deviation is shown in the figure for 1138-Bus when degree centrality was used as the method of node disruption. Normalised cost for spare capacity of 1 (closed squares), 2 (closed
circles), 3 (closed triangles), 4 (closed inverted triangles) and 5
(closed diamonds) percent.}
\label{fig:1138-Bus-DC-DD}
\end{figure}


\end{document}